\documentclass[aps,showpacs,a4paper,floatfix,twocolumn,prc,amsmath,amssymb]{revtex4-1}
\usepackage{graphicx}
\usepackage{epsfig}
\usepackage{ulem}
\usepackage{morefloats}
\usepackage{rotating}
\usepackage{relsize}

\newcommand{\pc}[1]{\ensuremath{\left(#1\right)}}

\newcommand{\lm}[1]{\ensuremath{\mathlarger #1}}

\begin{document}

\title{Light clusters, pasta phases and phase transitions in core-collapse supernova matter}

\author{Helena Pais, Silvia Chiacchiera, Constan\c ca Provid\^encia}
\affiliation{CFisUC, Department of Physics, University of Coimbra, P-3004-516 Coimbra, Portugal}

\begin{abstract}
The pasta phase in core-collapse supernova matter (finite temperatures and fixed proton fractions) is studied within relativistic mean field models. Three different calculations
are used for comparison, the Thomas-Fermi (TF), the Coexisting Phases (CP) and the
Compressible Liquid Drop (CLD) approximations. The effects of including light clusters in nuclear matter and the densities at which the transitions between pasta configurations and to uniform matter occur are also investigated. The free energy, pressure, entropy and chemical potentials in the range of particle number densities and temperatures expected to cover the pasta region are calculated. Finally, a comparison with a finite temperature Skyrme-Hartree-Fock calculation is drawn.
\end{abstract}
 
\pacs{24.10.Jv,26.50.+x,26.60.-c,26.60.Gj}

\maketitle

\section{Introduction} \label{I}

The complex structure of nuclear matter in the density region approaching $\rho_s \sim 0.16$ fm$^{-3}$  (central density of heavy nuclei) at finite temperature ($T < 20$ MeV) critically affects many astrophysical and nuclear physics phenomena. At low densities, the frustrated system called ``pasta phase'', caused by the competition between the Coulomb interaction and the strong force, appears, and is constituted by different geometrical configurations, as the density increases \cite{Ravenhall-83, Horowitz-04I, Horowitz-05, Maruyama-05, Watanabe-05a, Sonoda-08, Sonoda-10, Pais-12, Grill-14}.

The main interest in the pasta phase in core-collapse supernovae (CCSN) is that the neutrino opacity, which plays the main role in the development of a shock wave during the supernova collapse, is affected by its presence \cite{Horowitz-04I, Horowitz-04II, Sonoda-07}.
At very low densities (up to 0.001 times the saturation density), light nuclei (deuterons, tritons, helions, $\alpha$-particles) can appear \cite{Typel-10, Horowitz06, Heckel09, Ferreira12, Avancini-10Cl, Avancini-12Cl}, and, like the pasta phase, can modify the neutrino transport, which will have consequences in the cooling of the proto-neutron star \cite{Lattimer-91, Shen-98}.

Following our previous works (see e.g. \citep{Grill-14} and references therein), we use the Thomas-Fermi (TF) approximation, where the surface effects are treated self-consistently, in the framework of relativistic mean field (RMF) models, to calculate the pasta phase for a fixed proton fraction, several temperatures and densities. Besides TF, we use the coexisting-phases (CP) method, which is numerically much faster than the TF one, and where the Gibbs equilibrium conditions are used to get the lowest free energy state, and the surface and Coulomb terms are added ``by hand'' (see e.g. \citep{Avancini-12Cl}). Within this method, we also include the effect of light clusters. The compressible liquid drop (CLD) model is also considered. Unlike the CP approximation, this method takes into account both the Coulomb and surface terms in the minimization of the total energy of the system.

In the present work we perform a thermodynamical study of the phase transition between different shapes and we calculate the phase transformations occurring in the inner crust. We are interested in understanding the character of the crust-core transition and, for that reason, we study how the subsaturation instabilities are lifted by an appropriate description of the inner crust. In particular, we want to identify the strong and weak points of each approach. 

We refer to phase transitions and phase transformations as in Ref.~\cite{Ducoin06}, where the authors have stressed that there should be a distinction between ``continuous transition'' and ``continuous transformation''. They make the difference between a transformation, which is a specific path in the space of thermodynamic variables, and a phase transition, which is an anomaly of the thermodynamic potential considered in the total space of thermodynamic variables. Having this in mind, an asymmetric system, containing a fixed proton fraction, is going through a first-order phase transition, since the first derivatives of the grand potential are discontinuous, though the thermodynamic transformations may result in a continuous evolution of the observables.

We compare our results, obtained within the three specified methods, with the work of Raduta and Gulminelli \citep{Raduta10}, Hempel and Schaffner-Bielich \citep{Hempel-10}, Zhang and Shen \citep{Zhang14}, and Pais and Stone \cite{Pais-12}. Raduta et al. developed a phenomenological statistical model for dilute star matter, in which free nucleons are treated within a mean-field approximation, and nuclei are considered to form a loosely interacting cluster gas, with $T = 1 - 20$ MeV, $y_p= \rho_p/\rho = 0 - 0.5$ and $\rho > 10^8$ g/cm$^3$ ($\rho \gtrsim 6.02 \times 10^{-8}$ fm $^{-3}$), making it appropriate for CCSN description. They found that, for all sub-saturation densities, matter can be viewed as a continuous fluid mixture between free nucleons and massive nuclei. As a consequence, the equations of state (EoS) and the associated observables do not present any discontinuity over the whole thermodynamic range and the expected first-order transition to uniform matter does not happen. 
Hempel et al. also used a statistical model for the EoS that can be applied to all densities relevant for supernova simulations, where an ensemble of nuclei and interacting nucleons are in statistical equilibrium. On the other hand, Zhang and Shen studied the non-uniform matter using a self-consistent TF approximation, though they only considered two geometrical configurations, droplets and bubbles. They have compared their results with a parametrized TF calculation, where the surface energy and the nucleon distribution are calculated differently, and they have reached the conclusion that the parametrized approximation is a reasonable one. 
We also compare our results with a 3D finite temperature Skyrme-Hartree-Fock calculation \citep{Pais-12,Pais-14TN}, where four different Skyrme interactions have been used, and where subtle variations in the low and high density transitions into and out of the pasta phase were found.

The paper is organized as follows.
In section \ref{II}, we briefly review the formalism used and in section \ref{III}, the results are discussed. Finally, in section \ref{IV}, some conclusions are drawn.

\section{Formalism} \label{II}
 
We consider a system of baryons, with mass $M$ interacting with
and through an isoscalar-scalar field $\phi$ with mass $m_s$, an
isoscalar-vector field $V^{\mu}$ with mass $m_v$ and an
isovector-vector field $\mathbf b^{\mu}$ with mass $m_\rho$. When
describing $npe$ matter we also include a system of electrons with
mass $m_e$. Protons and electrons interact through the
electromagnetic field $A^{\mu}$. The Lagrangian density reads:

\begin{equation*}
    \mathcal{L}=\sum_{i=p,n}{\mathcal{L}_{i}} +\mathcal{L}_{e } 
    + \mathcal{L}_{{\sigma }}+ \mathcal{L}_{{\omega }} + 
\mathcal{L}_{{\rho }}+{\mathcal{L}}_{\gamma } ,
\end{equation*}

where the nucleon Lagrangian reads

\begin{equation}
\mathcal{L}_{i}=\bar{\psi}_{i}\left[ \gamma _{\mu }iD^{\mu }-M^{*}\right]
\psi _{i}  \label{lagnucl},
\end{equation}

with
\begin{eqnarray}
iD^{\mu } &=&i\partial ^{\mu }-g_vV^{\mu }-\frac{g_{\rho }}{2}{\boldsymbol{\tau}}%
\cdot \boldsymbol{b}^{\mu } - e \frac{1+\tau_3}{2}A^{\mu}, \label{Dmu} \\
M^{*} &=&M-g_s\phi
\end{eqnarray}
and the electron Lagrangian is given by
\begin{equation}
\mathcal{L}_e=\bar \psi_e\left[\gamma_\mu\left(i\partial^{\mu} + e A^{\mu}\right)
-m_e\right]\psi_e.
\label{lage}
\end{equation}

The isoscalar part is associated with the scalar sigma ($\sigma$)
field $\phi$, and the vector omega ($\omega$) field $V_{\mu}$,
whereas the isospin dependence comes from  the isovector-vector
rho ($\rho$) field $b_\mu^i$ (where $\mu$ stands for the four
dimensional space-time indices  and $i$ the three-dimensional
isospin direction index). The associated Lagrangians are:

\begin{eqnarray*}
\mathcal{L}_{{\sigma }} &=&\frac{1}{2}\left( \partial _{\mu }\phi \partial %
^{\mu }\phi -m_{s}^{2}\phi ^{2} -\frac{1}{3}\kappa \phi ^{3}-\frac{1}{12}%
\lambda \phi ^{4} \right) \\ 
\mathcal{L}_{{\omega }} &=&-\frac{1}{4}\Omega _{\mu \nu }\Omega ^{\mu \nu }+ \frac{1}{2}m_v^{2}V_{\mu }V^{\mu }+\frac{1}{4!}\xi g_{v}^{4}(V_{\mu}V^{\mu })^{2} \\
\mathcal{L}_{{\rho }} &=&-\frac{1}{4} \mathbf{B}_{\mu \nu }\cdot \mathbf{B}^{\mu\nu }+ \frac{1}{2} m_{\rho }^{2}\boldsymbol{b}_{\mu }\cdot \boldsymbol{b}^{\mu }\\
\mathcal{L}_{{\gamma }} &=&-\frac{1}{4}F _{\mu \nu }F^{\mu
  \nu }\\
\end{eqnarray*}
where $\Omega _{\mu \nu }=\partial _{\mu }V_{\nu }-\partial
_{\nu }V_{\mu }$, $\mathbf{B}_{\mu \nu }=\partial _{\mu
}\boldsymbol{b}_{\nu }-\partial _{\nu }\boldsymbol{b}_{\mu
}-g_{\rho }(\boldsymbol{b}_{\mu }\times
\boldsymbol{b}_{\nu })$ and $F_{\mu \nu }=\partial _{\mu
}A_{\nu }-\partial _{\nu }A_{\mu }$. 

The model comprises the following parameters:
three coupling constants $g_s$, $g_v$ and $g_{\rho}$ of the mesons
to the nucleons, the bare nucleon mass $M$, the electron mass
$m_e$, the masses of the mesons, the electromagnetic coupling
constant $e=\sqrt{4 \pi/137}$ and the self-interacting coupling
constants $\kappa$, $\lambda$ and $\xi$. In this Lagrangian
density, $\boldsymbol \tau$ is the isospin operator.

We use the FSU parametrization \cite{Todd05}, expected to describe well the crust \cite{Grill-14}, even if it does not describe a 2 $M_\odot$ neutron star. 
This parametrization also includes a nonlinear $\omega\rho$ coupling term, which affects the density dependence of the symmetry energy. This term is given by:
\begin{equation}
{\cal L}_{\omega \rho}=\Lambda_{v} g_v^2 g_\rho^2 \mathbf
b_{\mu}\cdot \mathbf b^{\mu}\, V_{\mu}V^{\mu}.
\end{equation}

The state that minimizes the energy of asymmetric nuclear matter is characterized by the distribution functions, $f_{0k\pm}$, of particles ($+$) and antiparticles ($-$) $k=p,n,e$, given by: 

\begin{equation}
f_{0 j \pm}= \frac{1}{1+e^{(\epsilon_{0 j} \mp \nu_j)/T}}, \quad j=p,n
\end{equation}
with
\begin{equation}
\epsilon_{0 j}=\sqrt{p^2+{M^*}^2},  \quad \nu_j=\mu_j - g_v V_0^{(0)}  - \frac{g_\rho}{2}\, \tau_j b_0^{(0)}  \label{chempot}
\end{equation}
and
\begin{equation}
f_{0 e \pm}= \frac{1}{1+e^{(\epsilon_{0 e} \mp \mu_e)/T}}, 
\end{equation}
with
\begin{equation}
\epsilon_{0e}=\sqrt{p^2+m_e^2},
\end{equation}
where $\mu_k$ is the chemical potential of particle $k=p,n,e$.

In the mean field approximation, the thermodynamic quantities of interest are given in terms
of the meson fields, which are replaced by their constant expectation values. For homogeneous neutral nuclear matter, the energy density, the entropy density, the free energy density, and the pressure are given, respectively, by \cite{Avancini-10,*Avancini-12,Avancini-08,Cavagnoli11}:

\begin{eqnarray}
\lm\varepsilon&=&\frac{1}{\pi^2}\sum_{j=p,n}\int dp\,p^2\epsilon_{0j}\left(f_{0j+}+f_{0j-}\right) \nonumber \\
&+&\frac{m_v^2}{2}V_0^2+\frac{\xi g_v^4}{8}V_0^4+\frac{m_\rho^2}{2}b_0^2+\frac{m_s^2}{2}\phi_0^2 \nonumber \\
&+&\frac{k}{6}\phi_0^3+\frac{\lambda}{24}\phi_0^4+3\Lambda g_\rho^2g_v^2V_0^2b_0^2, \label{energy} \\
\mathcal{S}&=&-\frac{1}{\pi^2}\sum_{j=p,n}\int dp\,p^2 \left[f_{0j+}\ln f_{0j+} \right. \nonumber \\
&+&(1-f_{0j+})\ln(1-f_{0j+})+ f_{0j-}\ln f_{0j-} \nonumber \\
&+&\left.(1-f_{0j-})\ln(1-f_{0j-})\right], \label{entropy}\\
\mathcal{F}&=&\lm\varepsilon-TS,\label{free} \\
P&=&\frac{1}{3\pi^2}\sum_{j=p,n}\int dp\frac{p^4}{\epsilon_{0j}}\left(f_{0j+}+f_{0j-}\right) \nonumber \\
&+&\frac{m_v^2}{2}V_0^2+\frac{\xi g_v^4}{24}V_0^4+\frac{m_\rho^2}{2}b_0^2-\frac{m_s^2}{2}\phi_0^2 \nonumber \\
&-&\frac{k}{6}\phi_0^3-\frac{\lambda}{24}\phi_0^4+\Lambda g_\rho^2g_v^2V_0^2b_0^2. \label{pressure} 
\end{eqnarray}

For the electrons, we have
\begin{eqnarray}
\lm\varepsilon_e&=&\frac{1}{\pi^2}\int dp\, p^2  \epsilon_{0e}\left(f_{0e+}+f_{0e-}\right), \label{energy_e}\\
\mathcal{S}_e&=&-\frac{1}{\pi^2}\int dp\,p^2 \left[f_{0e+}\ln f_{0e+}+(1-f_{0e+})\ln(1-f_{0e+})\right. \nonumber \\
&+&\left. f_{0e-}\ln f_{0e-}+(1-f_{0e-})\ln(1-f_{0e-})\right], \\
\mathcal{F}_e&=&\lm\varepsilon_e-TS_e, \label{free_e}\\
P_e&=&\frac{1}{3\pi^2}\int dp \frac{p^4}{\epsilon_{0e}}\left(f_{0e+}+f_{0e-}\right). \label{pressure_e}
\end{eqnarray}

We consider matter with fixed proton fraction that is neutrino free, and hence the neutrino pressure and energy density are zero \cite{Avancini-10}.

\subsection{Light clusters}

We use the same prescription as in \cite{Avancini-12Cl} to include light clusters ($d\equiv^2$H, $t\equiv^3$H, $\alpha\equiv^4$He, $h\equiv^3$He) in the model. The Lagrangian density becomes
\begin{equation}
    \mathcal{L}=\sum_{i=p,n,t,h}{\mathcal{L}_{i}} +\mathcal{L}_{\alpha} + \mathcal{L}_d
    +\mathcal{L}_{e } + \mathcal{L}_{{\sigma }}+ \mathcal{L}_{{\omega }} + 
\mathcal{L}_{{\rho }}+{\mathcal{L}}_{\gamma }.
\end{equation}
 
The $\alpha$ particles and the deuterons are described as in \cite{Typel-10}:
\begin{eqnarray}
\mathcal{L}_{\alpha }&=&\frac{1}{2} (i D^{\mu}_{\alpha} \phi_{\alpha})^*(i D_{\mu \alpha} \phi_{\alpha})-\frac{1}{2}\phi_{\alpha}^* \pc{M_{\alpha}^*}^2
\phi_{\alpha},\\
\mathcal{L}_{d}&=&\frac{1}{4} (i D^{\mu}_{d} \phi^{\nu}_{d}-
i D^{\nu}_{d} \phi^{\mu}_{d})^* \nonumber
(i D_{d\mu} \phi_{d\nu}-i D_{d\nu} \phi_{d\mu})\\
&-&\frac{1}{2}\phi^{\mu *}_{d} \pc{M_{d}^*}^2 \phi_{d\mu},
\end{eqnarray}
$\mathcal{L}_i$ is defined in Eq.~(\ref{lagnucl}) and, for all clusters, we have
\begin{eqnarray}
iD^{\mu }_j &=&i\partial ^{\mu }-g_{vj}V^{\mu }-\frac{g_{\rho j}}{2}{\boldsymbol{\tau}}%
\cdot \boldsymbol{b}^{\mu } - e \frac{1+\tau_3}{2}A^{\mu}, \nonumber \\
j&=&t,h,\alpha,d .
\end{eqnarray}

The effective masses of the clusters are given by
\begin{eqnarray}
M_t^*&=&3M-B_t \\
M_h^*&=&3M-B_h \\
M_\alpha^*&=&4M-B_\alpha \\
M_d^*&=&2M-B_d 
\end{eqnarray} 
with their binding energies being $B_t=8.482$MeV, $B_h=7.718$MeV, $B_\alpha=28.296$ MeV, and $B_d=2.224$ MeV. The fraction of a cluster, $Y_i$, is given by $Y_i=\rho_i/\rho, i=\alpha,h,d,t$.  For the coupling constants, we consider $g_{vj}=A_jg_v$ and $g_{\rho j}=|Z_j-N_j|g_\rho$, where $A_j$ is the mass number, and $Z_j, N_j$ are the proton and neutron numbers, respectively \cite{Avancini-12Cl}. The chemical potential of a cluster $j$ is defined as $\mu_j=N_j \mu_n + Z_j\mu_p$. More realistic parametrizations for the couplings of the light clusters have been proposed in \cite{Typel-10, Ferreira12}, which should be implemented.

\subsection{Thomas-Fermi and Coexisting Phases approximations}

We use the Thomas-Fermi (TF) approximation to describe the nonuniform $npe$ matter inside the Wigner-Seitz unit cell,  that is taken to be a sphere, a cylinder or a slab in three, two, and one dimensions  \cite{Avancini-10,Avancini-12,Grill-14}. In this approximation, $npe$ matter is assumed locally homogeneous and at each point
its density is determined by the corresponding local Fermi momenta. In 3D we consider spherical symmetry, in 2D we assume axial symmetry around the z axis, and in 1D reflexion symmetry is imposed. In the TF approximation, fields are assumed to vary slowly so that baryons can be treated as moving in locally constant fields at 
each point \cite{Maruyama-05}. In this approximation, the surface effects are treated self-consistently. Quantities such as the energy and entropy densities are averaged over the cells. The free energy and pressure are calculated from these two thermodynamical functions, using the usual expressions, see e.g. Ref. \cite{Avancini-08}.
 
In the Coexisting Phases (CP) method, matter is organized into separated regions of higher and lower density, the higher ones being the pasta phases, and the lower ones a background nucleon gas. The interface between these regions is sharp. Finite size effects are taken into account by a surface and a Coulomb terms in the energy density \cite{Avancini-12Cl}. 

By minimizing the sum $\lm\varepsilon_{surf}+\lm\varepsilon_{Coul}$ with respect to the size of the droplet/bubble, rod/tube, or slab one gets \cite{Ravenhall-83}
\begin{eqnarray}
\lm\varepsilon_{surf}&=& 2\lm\varepsilon_{Coul}, \label{surf} 
\end{eqnarray}
with
\begin{eqnarray}
\lm\varepsilon_{Coul}&=&\frac{2\alpha}{4^{2/3}}\left(e^2\pi\Phi\right)^{1/3}\left[\sigma D(\rho_p^I-\rho_p^{II})\right]^{2/3}, \label{coul}
\end{eqnarray}
where $\alpha=f$ for droplets, rods, slabs and $\alpha=1-f$ for tubes and bubbles, $f$ is the volume fraction of phase $I$, $\sigma$ is the surface energy coefficient and $\Phi$ is given by
\begin{eqnarray}
\Phi=\left\lbrace\begin{array}{c}
\left(\frac{2-D \alpha^{1-2/D}}{D-2}+\alpha\right)\frac{1}{D+2}, D=1,3 \\
\frac{\alpha-1-\ln \alpha}{D+2}, D=2 \quad .
\end{array} \right. 
\end{eqnarray}

The Gibbs equilibrium conditions are imposed to get the lowest energy state, and, for a temperature $T=T^I=T^{II}$, are written as 
\begin{eqnarray}
\mu_n^I&=&\mu_n^{II},  \label{cp} \\
\mu_p^I&=&\mu_p^{II}, \nonumber \\
P^I&=&P^{II}, \nonumber 
\end{eqnarray}
 where $I$ and $II$ label the high- and low-density phases, respectively. When clusters are present, there are equilibrium conditions for them too \cite{Avancini-12Cl}.
 
The total free energy density, and the total proton density of the system are given by
\begin{eqnarray}
F&=&fF^I + (1-f)F^{II} + F_e +\lm\varepsilon_{surf} + \lm\varepsilon_{Coul}, \label{totalfree}\\
\lm\rho_p&=&\lm\rho_e=y_p\lm\rho=f\lm\rho_p^I+(1-f)\lm\rho_p^{II} ,
\end{eqnarray}
where $F^i, i=I,II,$ is the free energy density of the homogeneous neutral nuclear matter, given by Eq.~(\ref{free}), $F_e$ is given by Eq.~(\ref{free_e}), and $\lm\varepsilon_{surf}$ and $\lm\varepsilon_{Coul}$ are the surface and Coulomb energies, given by Eqs.~(\ref{surf}) and (\ref{coul}), respectively. 
 
\subsection{Compressible Liquid Drop model}
 
In the Compressible Liquid Drop model \cite{Baym-71,Lattimer-85,Lattimer-91,Bao14}, the equilibrium conditions of the system are derived from the minimization of the total free energy \cite{Baym-71}, including the surface and Coulomb terms. This minimization is done with respect to four variables: the size of the geometric configuration, $r_d$, which gives, just like in the CP case, Eq.~(\ref{surf}), the baryonic density in the high-density phase, $\rho^{I}$, the proton density in the high-density phase, $\rho_p^I$, and the volume fraction, $f$. The equilibrium conditions become:

\begin{eqnarray}
\lm\mu_n^I&=&\lm\mu_n^{II},   \\ 
\lm\mu_p^I&=&\lm\mu_p^{II}-\frac{\lm\varepsilon_{surf}}{f(1-f)(\lm\rho_p^I-\lm\rho_p^{II})}, \nonumber \\
P^{I}&=&P^{II}-\lm\varepsilon_{surf}\left(\frac{1}{2\alpha}+\frac{1}{2\Phi}\frac{\partial\Phi}{\partial f}-\frac{\lm\rho_p^{II}}{f(1-f)(\lm\rho_p^I-\lm\rho_p^{II})}\right). \nonumber
\end{eqnarray}

Note that there is an extra term in both the proton chemical potential and in the mechanical equilibrium conditions, as compared to the ones obtained in the CP approximation, Eqs. (\ref{cp}). These terms arise from the inclusion of the surface and Coulomb terms in the minimization of the total energy. The Coulomb repulsion induces an extra positive term while the surface tension reduces the cluster internal pressure.  

The total pressure, and the total proton chemical potential of the system are given by
\begin{eqnarray}
P_{tot}&=&\lm\mu_p\lm\rho_p+\lm\mu_n\lm\rho_n+\lm\mu_e\lm\rho_e-F, \label{pretot}\\
\lm\mu_p&=&f\lm\mu_p^I+(1-f)\lm\mu_p^{II}, 
\end{eqnarray}
where $F$ is the total free energy density, given by Eq.~(\ref{totalfree}), and $f$ is the volume fraction of phase $I$.

\section{Results} \label{III}

In the present section, we discuss how the nuclear  liquid-gas
instability, occurring at subsaturation densities for asymmetric
nuclear matter, is partially lifted by an adequate description of the
inner crust, allowing for the appearance of nonhomogeneous phases.  In
particular,  we will compare several physical quantities obtained
within a TF calculation, a CP approach, supposing a zero thickness
surface, and the CLD model, where finite size effects are included in
a consistent way,  with the corresponding quantities for homogeneous matter. We will also discuss the effect of the inclusion of light clusters in the calculation.

The free energy per particle, the pressure, the proton, neutron and baryonic chemical potentials, and the entropy per particle of the inner crust, obtained within the  approaches referred above, are plotted in the following figures as a function of density or chemical potential, 
for the FSU interaction, two temperatures $T = 4$ MeV and 8 MeV, and the proton fraction $y_p= \rho_p/\rho = 0.3$.  The results are shown  for homogeneous matter (red),  CP (blue) calculations with (dashed) and without (solid line) clusters, CLD (green solid line) and TF (points) calculations.

For reference and to help the discussion,  we show in Table \ref{tab1} the symmetric nuclear matter properties for all the models we are using in this study to compare with our calculations with the FSU interaction: another RMF parametrization, TW \cite{Typel-99}, with density-dependent couplings, and four Skyrme interactions, SkM* \cite{Bartel-82}, SLy4 \cite{Chabanat-98}, NRAPR \cite{Steiner-05}, and SQMC700 \cite{Guichon-06}, chosen based on their overall performance in modelling a wide variety of nuclear matter properties \cite{Dutra-12}.

\begin{table}[!htbp]
  \centering
  \caption{Symmetric nuclear matter properties at saturation density $\rho_0$ (energy per particle $B/A$, incompressibility $K$, symmetry energy $E_{sym}$ and symmetry energy slope $L$) for the FSU parametrization, and five other parameter sets for comparison. All the quantities are in MeV, except for $\rho_0$, given in fm$^{\rm {-3}}$.} 
  \begin{tabular}{c c c c c c}
    \hline
    \hline
	Model & $\rho_0$ & $B/A$ & $K$ & $E_{sym}$ & $L$   \\
    \hline
	  	\vspace*{0.2cm}
    	FSU & 0.148 & -16.3 & 230 & 32.6 & 60.5 \\
	TW & 0.15 & -16.3 & 240 & 33 & 55 \\
	NRAPR & 0.16 & -15.85 & 226 & 33 & 60 \\
	SQMC700 & 0.17 & -15.49 & 222 & 33 & 59  \\
	SkM* & 0.16 & -15.77 & 217 & 30 & 46 \\
	SLy4 & 0.16 & -15.97 & 230 & 32 & 46 \\
	\hline   
    \hline
  \end{tabular}
 \label{tab1}
\end{table} 

\begin{figure}[!htbp]
   \includegraphics[width=0.5\textwidth]{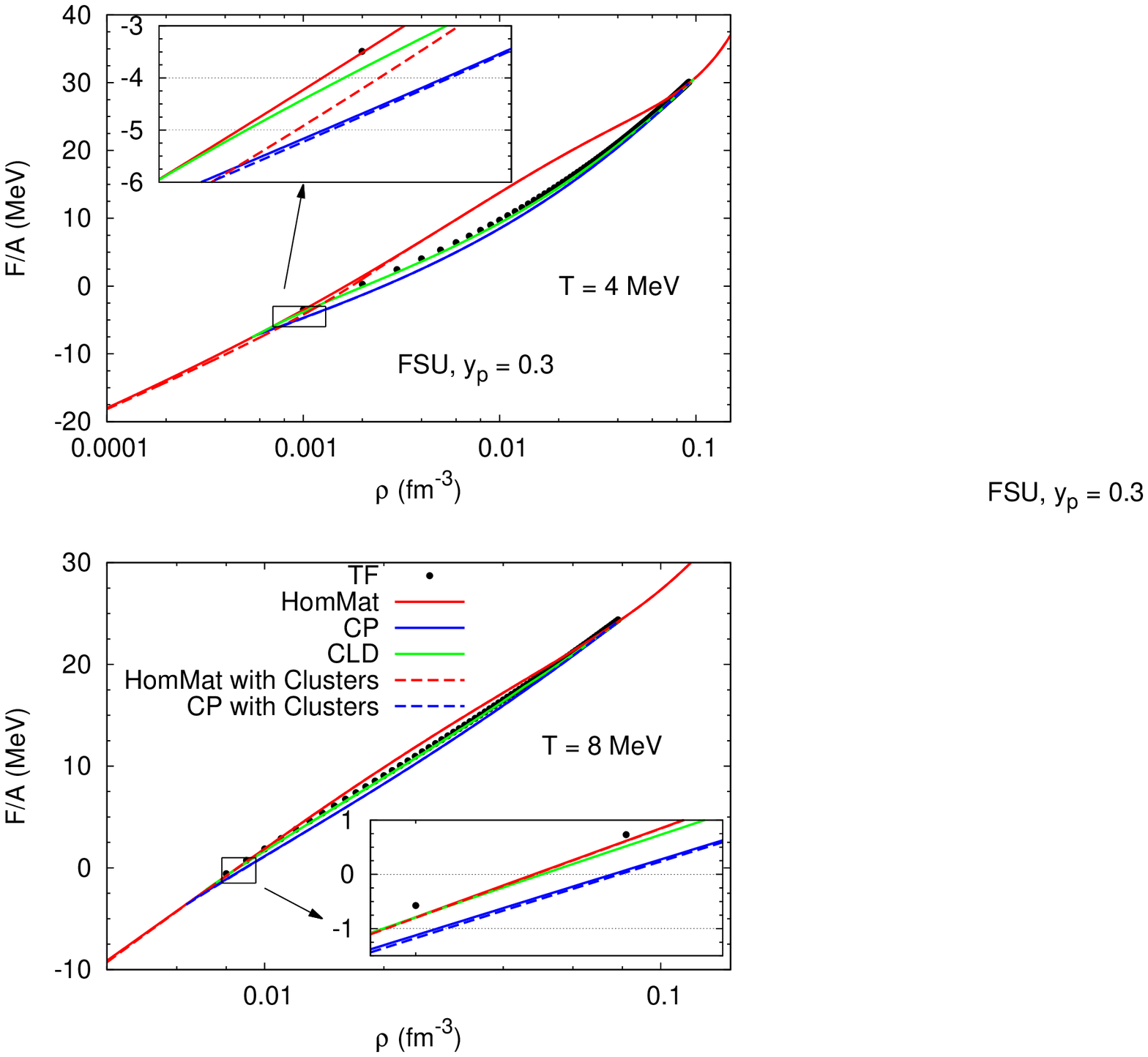} \\
  \caption{(Color online) Free energy per baryon as a function of the density for the FSU interaction, $y_p=0.3$, $T = 4$ MeV (top) and 8 MeV (bottom panel). Results with pasta (within CP, CLD and TF approaches) and for homogeneous matter, and including (for homogeneous matter and pasta within CP) or not clusters, are shown. The effect of these aggregates are only seen for very small densities (inset panels).}
\label{fig1}
\end{figure}

In Figure \ref{fig1}, we show the free energy per particle, $F/A$, as a function of the density.
As expected, $F/A$ is lowered when nonhomogeneous matter is present,
making these states more stable. A second effect is the disappearance
of the negative curvature that the EOS of homogeneous matter presents
below saturation density. This effect is present in all the three methods considered.

The light clusters are only present for very small densities, and will
start melting for $\rho\gtrsim 0.001$ fm$^{-3}$. However,  their
presence lowers the free energy of the homogenous matter EOS and of
the CP calculation, as can be seen in the inset panels. A TF
calculation, including light clusters, should also be performed (see
e.g. \cite{typel14}). The CP approach, which does not take into
account in a consistent way the surface tension and Coulomb energy,
overestimates the effect of the clusterization, mainly at low
densities. This problem is solved within the CLD approach, which, taking into
account  finite size effects in the phase equilibrium conditions,
gives results  closer to the TF calculation. However,  there is still
some overestimation of the free energy reduction with respect to the TF calculation, possibly due
to the approximate description of the surface energy. For the surface tension,
 we take a parametrization which is a function of the proton fraction and the
 temperature, and was fitted to the results obtained from a
 relativistic Thomas-Fermi calculation for a semi-infinite slab \cite{Avancini-12Cl}. 
 Close to the crust-core transition, all approaches, TF, CP and CLD,
 give similar results, and predict a first-order phase transition to
 uniform matter. A first order phase transition at the crust-core
 transition has also been obtained within other approaches \cite{Hempel-10,Pais-14TN}. 

When the temperature is increased to 8 MeV, similar conclusions are drawn, the main differences being a  decrease of the free energy and  the density range of the pasta. A remark is in order: the first two points of the free energy for the TF calculation are above the homogeneous matter value. This should be explained by the fact that, in the TF, we considered different values for the masses of the protons and neutrons, and by the fact that, for high temperatures, the precision in the calculation begins to be very critical. Let us, however, point out that 8 MeV is already a quite high temperature, and thermal fluctuations of the clusters rod-like or slab-like will destroy the Wigner-Seitz structures according to \cite{Pethick-98}.

\begin{figure}[!htbp]
   \includegraphics[width=0.5\textwidth]{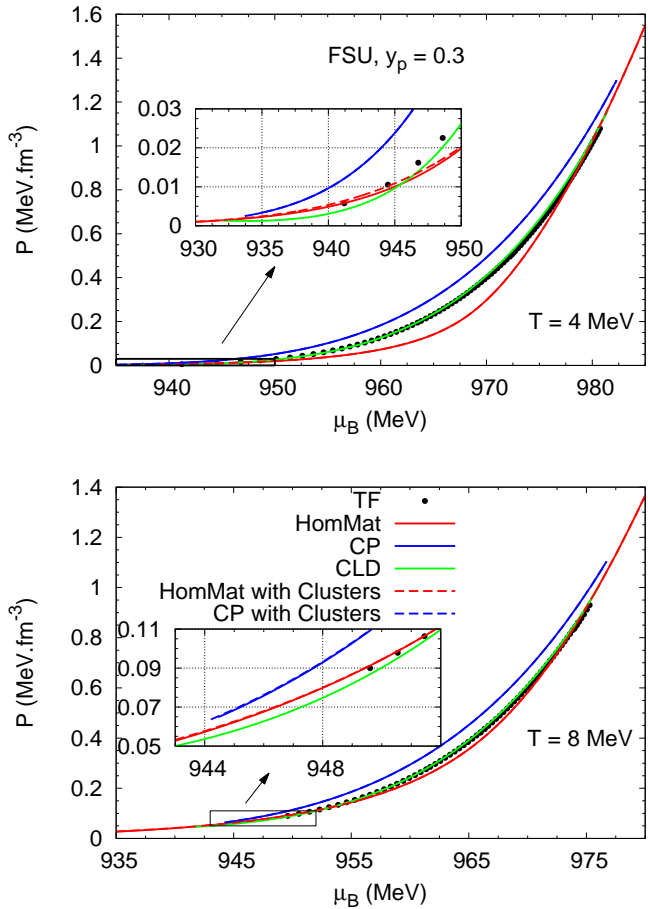} \\
   \caption{(Color online) Total pressure as a function of the baryonic chemical potential for the FSU interaction, $y_p=0.3$, $T = 4$ MeV (top) and 8 MeV (bottom panel). Results with pasta (within CP, CLD and TF approaches) and for homogeneous matter, and including (for homogeneous matter and pasta within CP) or not clusters, are shown. }%
\label{fig2}
\end{figure}

In Figure \ref{fig2}, the total pressure for the FSU interaction is plotted as a function of the baryonic chemical potential, $\mu_B$, which is defined as in Ref. \citep{Hempel-10}, 
\begin{equation}
\mu_B = (1-y_p)\mu_n+y_p(\mu_p+\mu_e),
\label{mub}
\end{equation} 
since we are performing a calculation with a fixed proton fraction and
charge neutrality. The pressure is a  smooth function of $\mu_B$ and,
within the CLD and TF approaches, does not show any discontinuity. 
The CP approach does not show any discontinuities between the  intermediate shapes, however, at  the onset of the pasta phase and at
  the crust-core transition, the pasta phase pressure does  not match smoothly into the homogenous matter pressure.  This is probably due to the non-consistent treatment of the surface energy.  The free energy of the pasta matches the homogeneous matter free energy with a different slope,  both at low densities and at the crust-core density,  giving rise to discontinuities in the pressure.
 The explicit inclusion of non-homogeneous matter, also true for the light clusters, increases the total pressure turning matter more stable. Consistently with the conclusions drawn for the total free energy, the contribution calculated within the CP approach is larger than the contribution evaluated within TF, showing the limitations of the first method. The CLD approach presents results very similar to the TF calculation. It is interesting to notice that the effect of light clusters is to increase the pressure, making the system more stable.

\begin{figure}[!htbp]
   \includegraphics[width=0.5\textwidth]{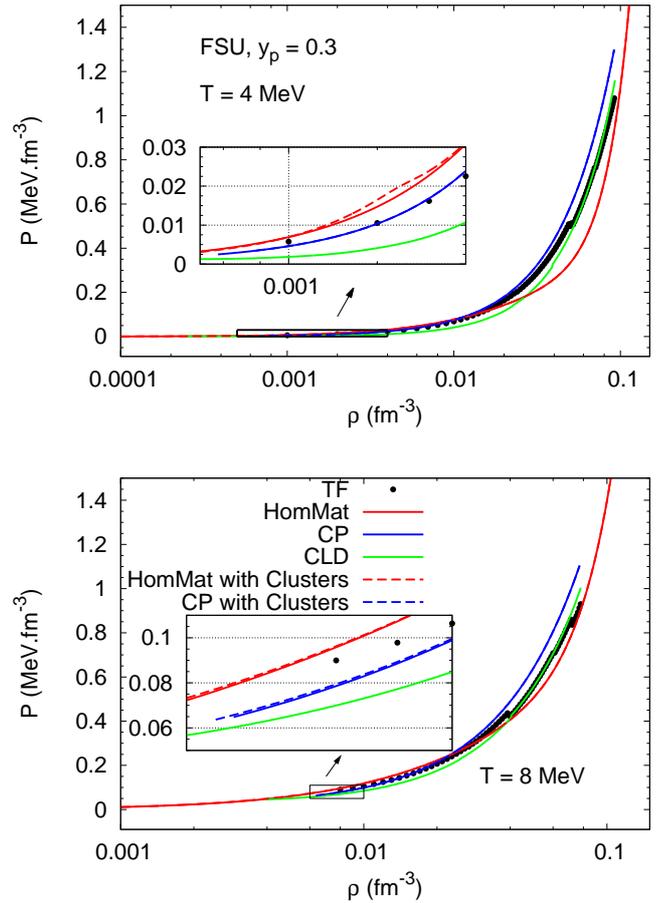} \\
   \caption{(Color online) Total pressure as a function of the density for the FSU interaction, $y_p = 0.3$, $T = 4$ (top) and 8 (bottom) MeV. Results with and without pasta, and including or not clusters, are shown. The effect of these aggregates are only seen for very small densities (inset panels).}
\label{fig3}
\end{figure}

Let us now analyse the behavior of the pressure with the baryonic density. In particular, we will discuss how the instability at subsaturation densities in asymmetric nuclear matter is lifted in stellar matter. 
In Figure \ref{fig3}, the total pressure is plotted against the
density, for the FSU interaction. First, it is interesting to notice
that there are no instabilities, that is, there is no range of densities
with a negative slope, except for small discontinuities, when the
transition from one shape to  another occurs, in the TF and CLD calculations. In the CP calculation, those discontinuities do not appear, because of the method itself: the Gibbs conditions, $P^I=P^{II}$, are imposed and no
contribution from the surface tension is taken into account. There is,
however, a jump at the onset and melting of the inner crust, precisely
due to the simplified description of the surface. 
 The occurrence of discontinuities between the different pasta phases
 in the TF and CLD calculations originates  from the simplified
 approach we have considered to describe the pasta phase, when only
 some geometries are included. However, in the TF calculation, even
 with this restriction, not always a transition between different
 shapes is discontinuous, e. g. the transition droplet--rod that
 occurs at $\rho_{d-r} = 0.024$ fm$^{-3}$, for $T = 4$ MeV, and at
 $\rho_{d-r} = 0.023$, for $T = 8$ MeV, is continuous. 

 At the crust core transition we observe a discontinuity in all methods, with a larger density jump for the CP method and a  smaller one for TF.  A similar behavior has been obtained in other
   approaches \cite{Lattimer-91,Hempel-10,Pais-12,okamoto2012}.
Since the pressure  does not change continuously into the homogeneous matter, a  Maxwell construction may be used  to describe the transition from the non-uniform to uniform matter, imposing a fixed proton fraction  and charge neutrality, and equal pressures and baryonic chemical potential as defined in Eq. (\ref{mub}), in both phases.
 Another aspect that should also be pointed out is that the onset of the pasta phases within the TF approach occurs at a larger density than the CLD one. The TF approach has also several limitations, and a calculation with the extended TF should
 be performed. However, we could also consider that, at smaller densities,
 the clustering onset is defined by the appearance of light clusters.

\begin{figure}[!htbp]
   \includegraphics[width=0.5\textwidth]{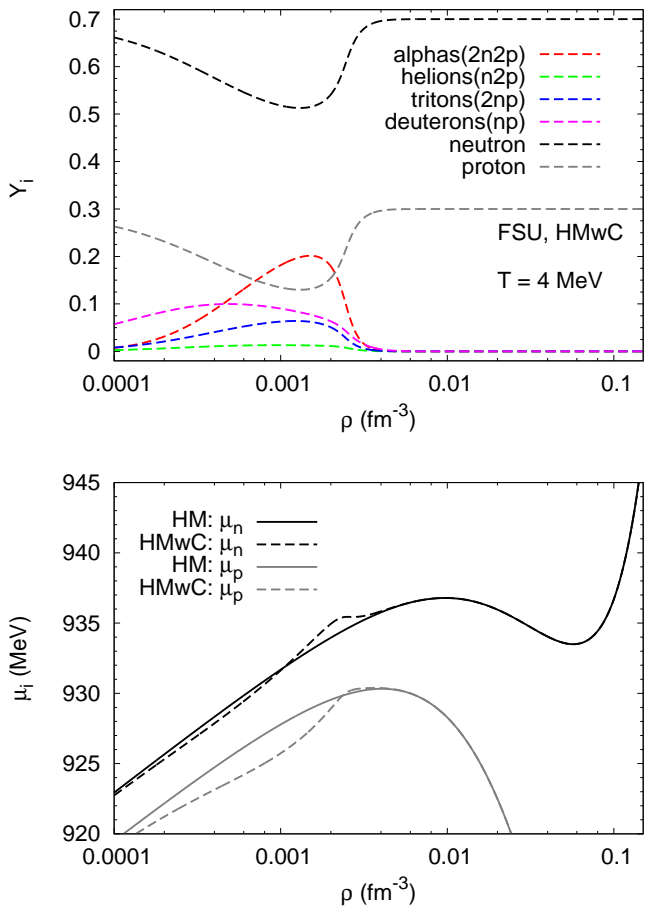} \\
\caption{(Color online) Fractions of the nucleons and clusters (top) and chemical potentials of the nucleons (bottom) as a function of the density for the FSU interaction, $y_p=0.3$ and $T = 4$ MeV, for homogeneous matter with clusters.}%
\label{fig4}
\end{figure}

\begin{figure}[!htbp]
   \includegraphics[width=0.5\textwidth]{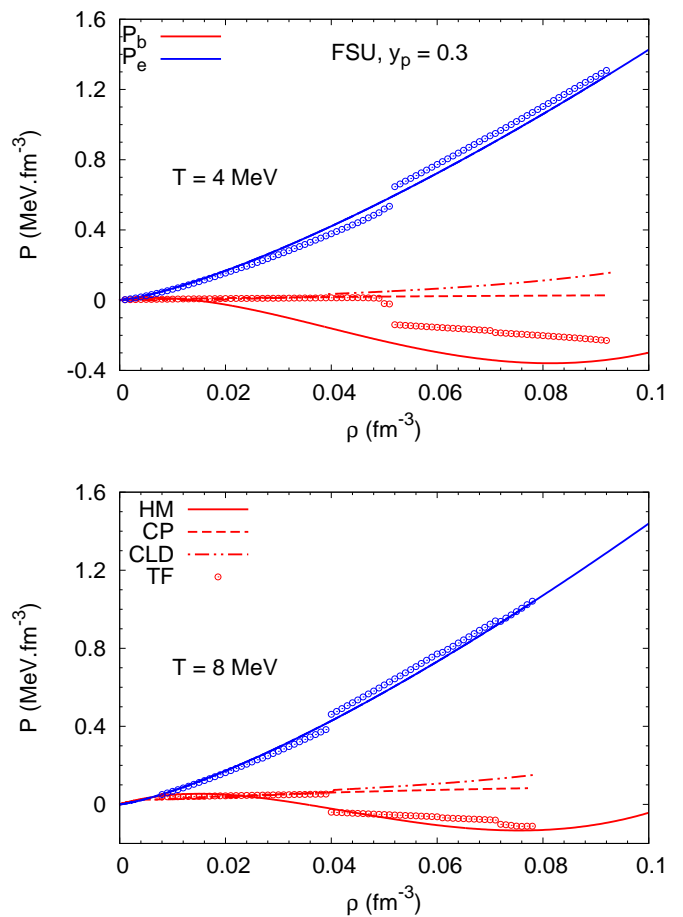} \\
  \caption{(Color online) Pressure as a function of the density for the FSU interaction, $y_p=0.3$, and $T = 4$ MeV (top) and 8 MeV (bottom panel). Homogeneous matter (HM) (solid), CP (dashed line), CLD (dash-point), and TF (points) results are shown for the baryonic (red), and electronic (blue) components. }
\label{fig5}
\end{figure}

We can also notice the effect of the clusters if we look at the inset
panels in both plots. We observe that, for $T=4$ MeV and $\rho < 0.001$ fm$^{-3}$, the clusters slightly lower the pressure, bringing the homogeneous matter result closer to the TF calculation. Above $\rho>0.001$ fm$^{-3}$, the clusters increase the pressure.
 Increasing the temperature, the clusters increase
 slightly the pressure in all range of densities shown, although for very low
 densities, the inclusion of light clusters still lowers the pressure. This occurs because the formation of light clusters increases the neutron fraction of the homogeneous matter,
 since light clusters are preferentially symmetric particles (deuterons or
 $\alpha$ particles) and, at high temperatures, the extra binding due to
 the cluster formation does not compensate the extra repulsion
 homogeneous matter with a larger neutron fraction experiences.  This is
 clearly seen in Fig. \ref{fig4}, where we plot, for $T = 4$ MeV, the
 neutron, proton, and light clusters fractions versus baryonic
 density in the top panel, and the neutron and proton chemical
 potentials in the bottom panel. An increase of the neutron chemical
 potential above the homogeneous matter one occurs close to the maximum
 of $\alpha$ clusters. This comes together with a decrease of the proton
 and neutron fractions, and a decrease of the proton fraction in the
 homogeneous matter. Let us point out, however, that the thermodynamical function we should look at to discuss the extra stability the light clusters give to the system is the free energy, when the baryonic density is taken as a state variable.

In order to understand the role of the electronic contribution to the total pressure,
we plot in Fig. \ref{fig5} the baryonic (red) and electronic
(blue) pressure components as a function of the density. This separation is not totally possible for the TF
calculation due to the coupling
between protons and electrons, induced by the Coulomb interaction. In the CLD method, and because of this separation problem, we are calculating $P_b$, given by Eq. (\ref{pressure}), as $P_b=fP^I+(1-f)P^{II}$, unlike the other pressure plots, where it is calculated from the thermodynamical expression, given by Eq.(\ref{pretot}). The electronic pressure, $P_e$, is given by Eq. (\ref{pressure_e}).
 Within the CP and CLD approaches, the electron contribution coincides
 with the homogeneous matter contribution. We see that for CP, the
 baryonic pressure is always positive because it corresponds to the pressure at the binodal surface, that is always positive. In the TF calculation, the baryonic pressure still exhibits a region with a negative incompressibility, unlike the total pressure shown in Fig. \ref{fig2}. In fact, within TF, the negative slope of the pressure for nuclear homogeneous matter is only partially removed with the inclusion of droplets and the pasta phases.  The electrons are responsible for making the total pressure positive, and with positive slope. It is also seen that the effect of clusterization is washed out with the temperature, so that, for $T=8$ MeV, the pressure for the pasta calculation does not differ much from the pressure for homogeneous matter. Similar results were discussed in Ref.~\cite{Raduta10}, where subsaturation densities are described within a statistical model, which considers a continuous fluid mixture of  free nucleons and finite nuclei, the only difference being the absence of discontinuities between shape transitions. 
The discontinuity occurring in the electron pressure, within the TF
calculation, is due to the normalization of the Coulomb field, which is set to zero at the Wigner-Seitz cell border. This gives rise to a jump going from slabs to tubes because, for the first geometry, the proton density is almost zero at $r=R_{WS}$, while for the tubes it takes the largest value.

\begin{figure}[!htbp]
   \includegraphics[width=0.5\textwidth]{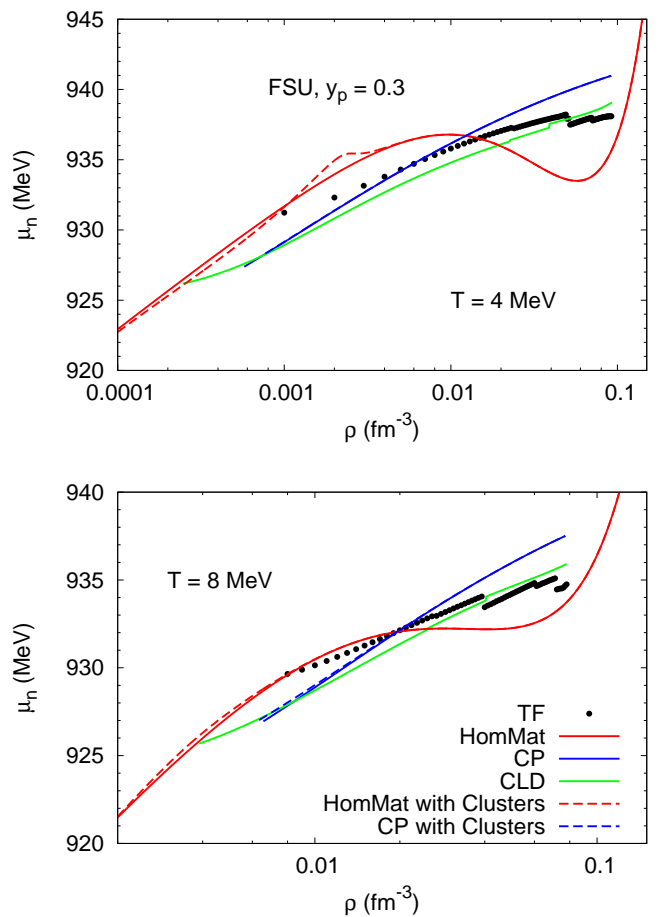} \\
   \caption{(Color online) Neutron chemical potential as a function of the density for the FSU interaction, $y_p=0.3$, and $T = 4$ MeV (top) and 8 MeV (bottom panels). Results with and without pasta, and including or not clusters, are shown. The effect of these aggregates are only seen below 0.01 fm$^{-3}$. }%
\label{fig6}
\end{figure}

\begin{figure}[!htbp]
   \includegraphics[width=0.5\textwidth]{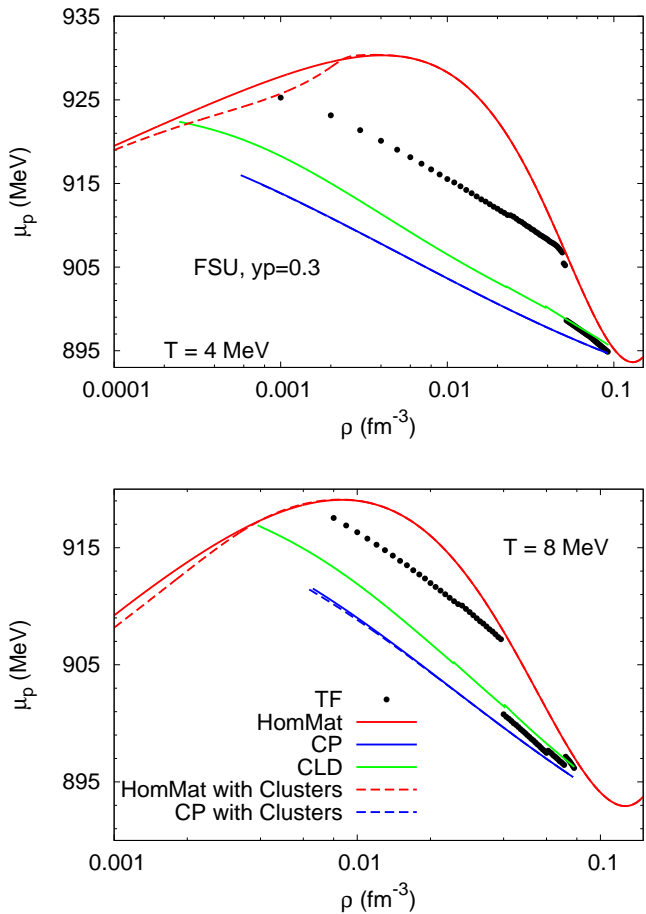} \\
   \caption{(Color online) Proton chemical potential as a function of the density for the FSU interaction, $y_p=0.3$, and $T = 4$ MeV (top) and 8 MeV (bottom panels). Results with and without pasta, and including or not clusters, are shown. The effect of these aggregates are only seen below 0.01 fm$^{-3}$. }
\label{fig7}
\end{figure}

In Figures \ref{fig6} and \ref{fig7}, the neutron and proton chemical potentials are plotted as a function of the density, for the FSU interaction. 
Both quantities show an instability for homogeneous matter,  explicitly  seen through the presence of a backbending: both chemical potentials decrease in a given range of densities, although the corresponding particle density is increasing. 

In Fig.~\ref{fig6},  we plot the neutron chemical potential. The
pasta phase practically removes the regions of instability. However,
discontinuities  are observed for all the intermediate shape
transitions in the TF and CLD calculations, and a large jump occurs at
the onset and melting of the pasta phase in the CP approach, and only
at the melting of the crust, for the TF and  CLD calculations.  It is
also seen that the light clusters are affecting the neutron chemical
potential, particularly for $T=4$ MeV: below $\rho=0.001$fm$^{-3}$, their
presence decreases the chemical potential, while above that density,
corresponding to the onset of the pasta phase, light clusters increase
$\mu_n$, as already discussed in Fig.~\ref{fig3}. This is due to the the appearance of more symmetric light
clusters at larger densities, that induces a more neutron rich
homogeneous matter.

\begin{figure}[!htbp]
   \includegraphics[width=0.5\textwidth]{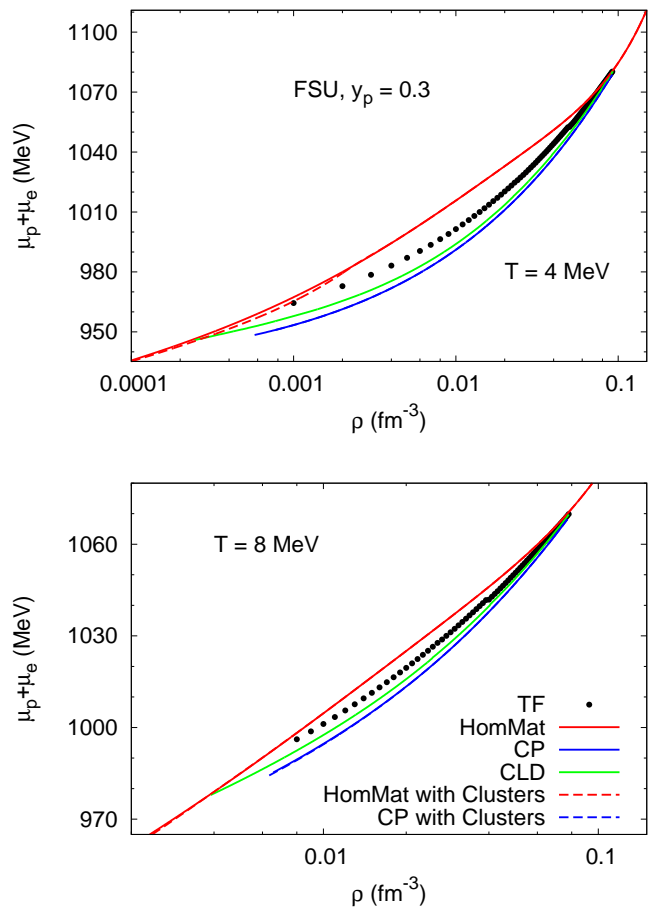} \\
  \caption{(Color online) The $\mu_p+\mu_e$ chemical potential as a function of the density for the FSU interaction, $y_p=0.3$, and $T = 4$ MeV (top) and 8 MeV (bottom). Results with and without pasta, and including or not clusters, are shown. The effect of these aggregates are only seen for very small densities.}%
\label{fig8}
\end{figure}

The proton chemical potential (Fig.~\ref{fig7}) behaves
differently: even though the back bending is reduced, it is still
clearly seen. 
In homogeneous matter, for a given  baryonic density,  the proton fraction is considered fixed and the charge neutrality is imposed. Consequently, there is only one global chemical potential associated to the global conservation of the baryonic number, Eq. (\ref{mub}), which is obtained from the derivative of the free energy with respect to the baryonic density.
 However, the calculation  of the pasta phases in the present discussion was done imposing a global proton fraction together with
 the charge neutrality condition. Therefore, the neutron density and proton densities are independent, since neutrons and protons can be
 exchanged freely between the two phases. Taking as degrees of freedom the baryon number and charge number, the neutron chemical potential
 is  $\mu_n=\mu_B$ and the proton chemical potential is $\mu_p=\mu_B+\mu_C$. In neutrino free matter, the leptonic chemical
 potential is zero and the electron chemical potential is $\mu_e=-\mu_C$. Putting together the proton and electron chemical
 potentials, we conclude that $\mu_p+\mu_e=\mu_B$, see also the  discussion in Ref. \cite{hempel09}.
In Fig.~\ref{fig8}, the chemical potential
$\mu_p+\mu_e$  is plotted and no backbending is observed  for the
pasta calculations, showing that the instability has been totally
raised, just like we have seen above for the neutron chemical
potential. The large jump occurring at the slab-tube transition in Fig.~\ref{fig7} is
related with the boundary conditions on the Coulomb field, when
integrating the equation of motion. Measurable physical quantities are
not affected by the boundary condition, but quantities, such as the chemical potential, are
particularly sensitive to the choice. 

\begin{figure}[!htbp]
   \includegraphics[width=0.5\textwidth]{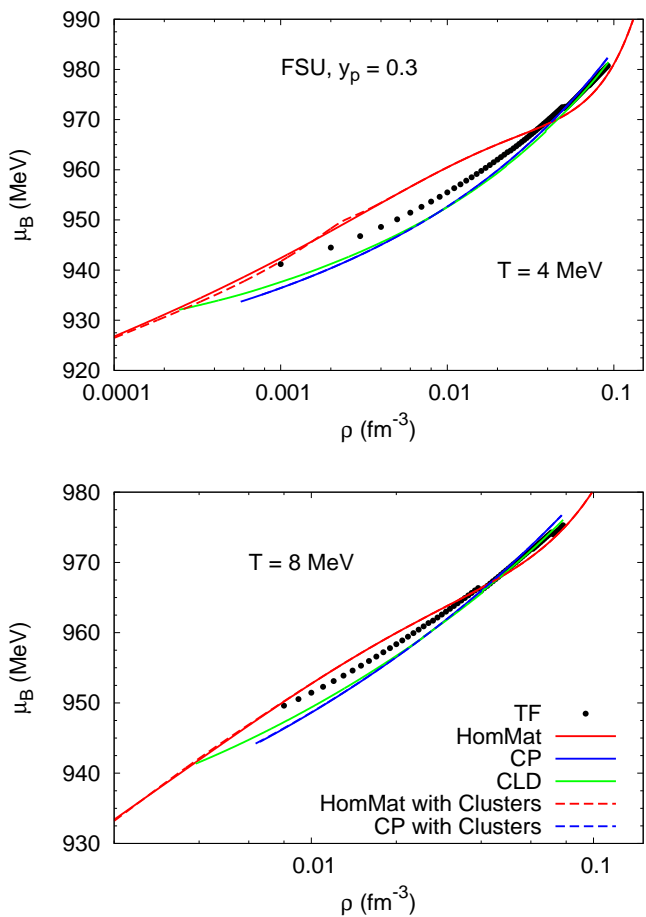} \\
  \caption{(Color online) Baryonic chemical potential as a function of the density for the FSU interaction, $y_p=0.3$, and $T = 4$ MeV (top) and 8 MeV (bottom). Results with and without pasta, and including or not clusters, are shown. The effect of these aggregates are only seen for very small densities.}%
\label{fig9}
\end{figure}

In Figure \ref{fig9}, we plot the baryonic chemical potential $\mu_B$, defined in Eq. (\ref{mub}), as a function of the baryonic density. For both temperatures shown, $\mu_B$ is a monotonically increasing function of the density, even for homogeneous matter. However, by including a phase of  non-homogeneous matter, the negative curvature of the chemical potential is removed.

At the crust-core transition, TF, CP, and CLD give similar
results. However, at the onset of the pasta phase, while TF and CLD
link continuously to homogeneous matter with clusters, the CP
calculation presents a very large discontinuity, reflecting the
non-consistent inclusion of the surface tension. This behavior is seen
in all chemical potential figures (except for $\mu_p$ at $T=8$ MeV for
the TF calculation). Light clusters reduce the homogeneous matter and
the  CP proton chemical potential, because clusters bring extra binding
to the system, in particular, the $\alpha$ particles (see Fig.~\ref{fig4}).

The discontinuities on the chemical potentials between the  different geometrical configurations are an indication of the
  limitations of the present approaches. A rearrangement of matter and
  charge will wash out  these discontinuities. Surface effects such as
  electrical double layers as the ones occurring on the boundary
  between charged solids and liquids, with an adsorption layer and a
  screening layer, would give rise to a continuous behavior of the
  charge chemical potential. Also the Wigner Seitz approximation may
  disfavor an optimal matter rearrangement. 
Quantum molecular dynamics calculations go beyond the Wigner Seitz
approximation and do not assume any specific non-uniform structure of baryon matter, but consider an uniform background of electrons \cite{Watanabe-05a, Sonoda-08, Sonoda-10}. 
Using a large enough cell to include several units of the pasta structures,  it has been shown recently in  \cite{okamoto2012,okamoto2013} that a fcc crystalline structure for the
 droplet phase would be favored with respect to the bcc one for
 densities just before the transition to a rod like phase. An important
 conclusion of both approaches was the natural appearance of the typical pasta phases with rod, slab, tube, and bubble, in addition to spherical droplets, when no assumption on the structures was used.

The behavior described above for both  the proton and neutron chemical potential is also obtained in \citep{Zhang14}, within a  self-consistent TF description of the pasta phase, including only two configurations, droplets and bubbles. The discontinuities  obtained are attributed to the Coulomb field.  Small discontinuities at low temperatures are obtained in  \citep{Hempel-10}, within a statistical model, due to a stepwise increase of the average number of nucleons and protons of the clusters with the density. Although the overall behavior is the same in  \citep{Raduta10}, where also a statistical model is applied, and inhomogeneities are described as a continuous mixture of loosely interacting clusters, the chemical potentials show no discontinuities. The smooth increase of the  average fragment mass fraction and cluster mass fraction explains this difference.

\begin{figure}[!htbp]
   \includegraphics[width=0.5\textwidth]{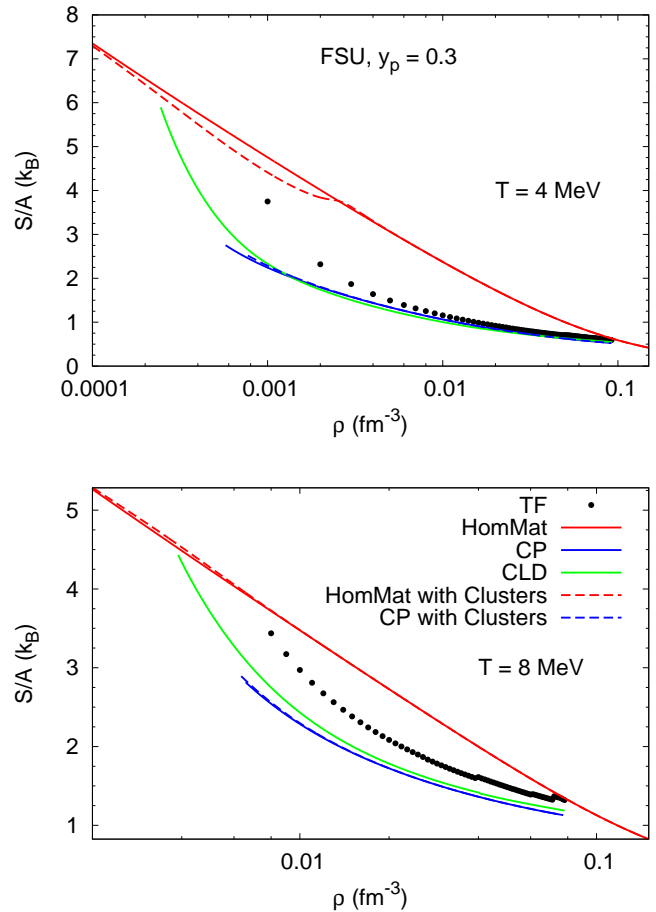} \\
  \caption{(Color online) Total entropy per baryon as a function of the density for the FSU interaction, $y_p=0.3$, and $T = 4$ MeV (top) and 8 MeV (bottom panels). Results with and without pasta, and including or not clusters, are shown. The effect of these aggregates are only seen for very small densities.} 
\label{fig10}
\end{figure}

In Fig.~\ref{fig10}, the total entropy per baryon is plotted as a function of the density.
The TF, CP and CLD calculations lower the entropy per particle due to the formation of heavy clusters. For $T=4$ MeV, and at low densities, the same effect is seen in homogeneous matter with light clusters. On the other hand, in the CP calculation, the inclusion of light clusters increases slightly the entropy because they reduce the formation of heavy clusters. When temperature is increased, the entropy for the pasta calculation gets closer to the homogeneous matter result because the  nucleons drip out of the heavy clusters.
The reduction of the entropy with clusterization was previously discussed in several works \cite{Shen-98,Hempel-10,Raduta10,Zhang14}.
 Looking at Fig. 16 of Ref. \citep{Hempel-10} and Fig. 21 of
 Ref.~\citep{Raduta10}, applying a statistical description, a smooth decrease of the entropy per baryon with the density is obtained. Similar results have been obtained in \citep{Zhang14}
 (see Fig. 2), within a self-consistent TF (STF) calculation for
 droplets and bubbles.

\begin{figure}
   \begin{tabular}{c}
   \includegraphics[width=0.5\textwidth]{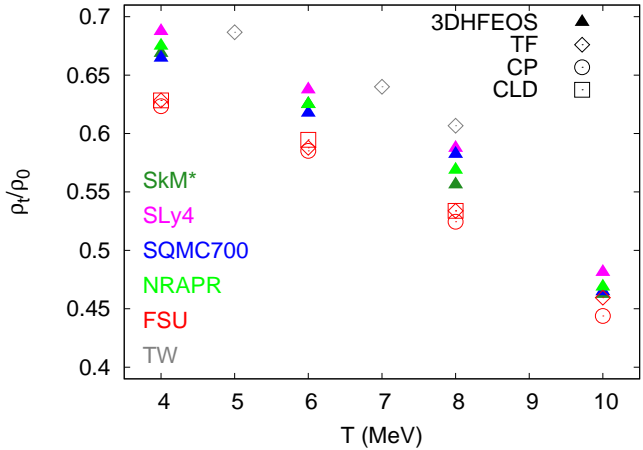} 
   \end{tabular}
\caption{(Color online) Crust-core transition densities, normalized to the nuclear saturation density $\rho_0$, as a function of the temperature, for the 3D-HFEOS (upward triangles) calculation and the SkM* (green), SLy4 (pink), SQMC700 (blue), NRAPR (yellow) models, and the Thomas-Fermi (diamonds) calculation for the FSU (red) and TW (grey) models, and the CP (circles), and CLD (squares) results for the FSU (red) model. The proton fraction is fixed to 0.3.}
\label{fig11}
\end{figure}

\begin{table*}
  \centering
    \caption{Transition densities in the pasta phase for the TF, CP and CLD calculations for the FSU interaction. $\rho_{HM}$ is the onset density of homogeneous matter. These values are the ones represented in Fig. \ref{fig12}.}\label{tab2}
  \begin{tabular}{l|ccccccccccccccccccc}
    \hline\hline
    $T$ & \multicolumn{4}{c}{ $\rho_{d-r}$ } & \multicolumn{4}{c}{ $\rho_{r-s}$ } & \multicolumn{4}{c}{ $\rho_{s-t}$ } & \multicolumn{4}{c}{ $\rho_{t-b}$ } & \multicolumn{3}{c}{ $\rho_{HM}$ } \\
(MeV) & \multicolumn{19}{c}{ $(\mbox{fm}^{-3})$}  \\
\hline
	&	TF & CP	& CLD & \phantom{aa} & TF & CP & CLD & \phantom{aa} & TF & CP & CLD & \phantom{aa} & TF & CP & CLD & \phantom{aa} & TF & CP & CLD   \\	
    4  & 0.024 & 0.032 & 0.023 & \phantom{aa} & 0.050 & 0.050 & 0.039 & \phantom{aa} & 0.052 & 0.085 & - & \phantom{aa} & 0.071 & - & - & \phantom{aa} & 0.093 & 0.092 & 0.093 \\
    6  & 0.025 & 0.032 & 0.024 & \phantom{aa} & 0.047 & 0.049 & 0.040 & \phantom{aa} & - & 0.083 & - & \phantom{aa} &0.068 & - & - & \phantom{aa} & 0.087 & 0.086 & 0.088 \\
    8  & 0.027 & 0.032 & 0.025 & \phantom{aa} & 0.040 & 0.048 & 0.040 & \phantom{aa} & - & - & - & \phantom{aa} & 0.061 & - & - & \phantom{aa} & 0.079 & 0.078 & 0.079 \\
    10 & 0.031 & 0.032 & - & \phantom{aa} & - & 0.047 & - & \phantom{aa} & - & - & - & \phantom{aa} & 0.047 & - & - & \phantom{aa} & 0.068 & 0.066 & - \\
    \hline\hline
  \end{tabular}
\end{table*}

In Fig.~\ref{fig11}, we show the crust-core transition densities for the
range of temperatures 4 to 10 MeV, and we compare them with the
results found for the transition densities to homogeneous matter with
a 3D  finite temperature  Skyrme-Hartree-Fock (3DHFEOS) calculation
\cite{Pais-12}, and with the results found in Ref.~\cite{Avancini-12}, for the TF calculation with the density-dependent TW interaction.  We see that the three approaches, TF, CLD and CP, give very similar results for the FSU interaction, and lower transition densities than the values found within a 3DHFEOS calculation, done in the framework of Skyrme interactions. The TW parameter set has a higher transition density between shapes, and to uniform matter, than the FSU interaction, and very close to the values found for the Skyrme forces. This might be explained by the fact that the behavior of the symmetry energy for RMF models with density-dependent couplings is much closer
to the behavior of Skyrme forces \cite{Ducoin-08}. It is known that the crust-core transition density  decreases with an increasing slope $L$ \cite{Vidana09,Ducoin11},
and, therefore, this may influence the results wtih  SkM$^*$ and SLy4,
having $L=46$ MeV, while for all the other models,  $L\sim 60$ MeV. Comparing
models with a similar $L$, we conclude that, within the 3DHFEOS, the
crust-core transition densities are  larger  by $\sim 0.04\rho_0$ than the transition densities for FSU, and this difference decreases with increasing temperature. The 3DHFEOS calculation allows a larger
freedom in the minimization of the free energy. However, differences
may also be due to the different energy density functionals generated
by each model.

\begin{figure}[!htbp]
   \begin{tabular}{c}
\includegraphics[width=0.5\textwidth]{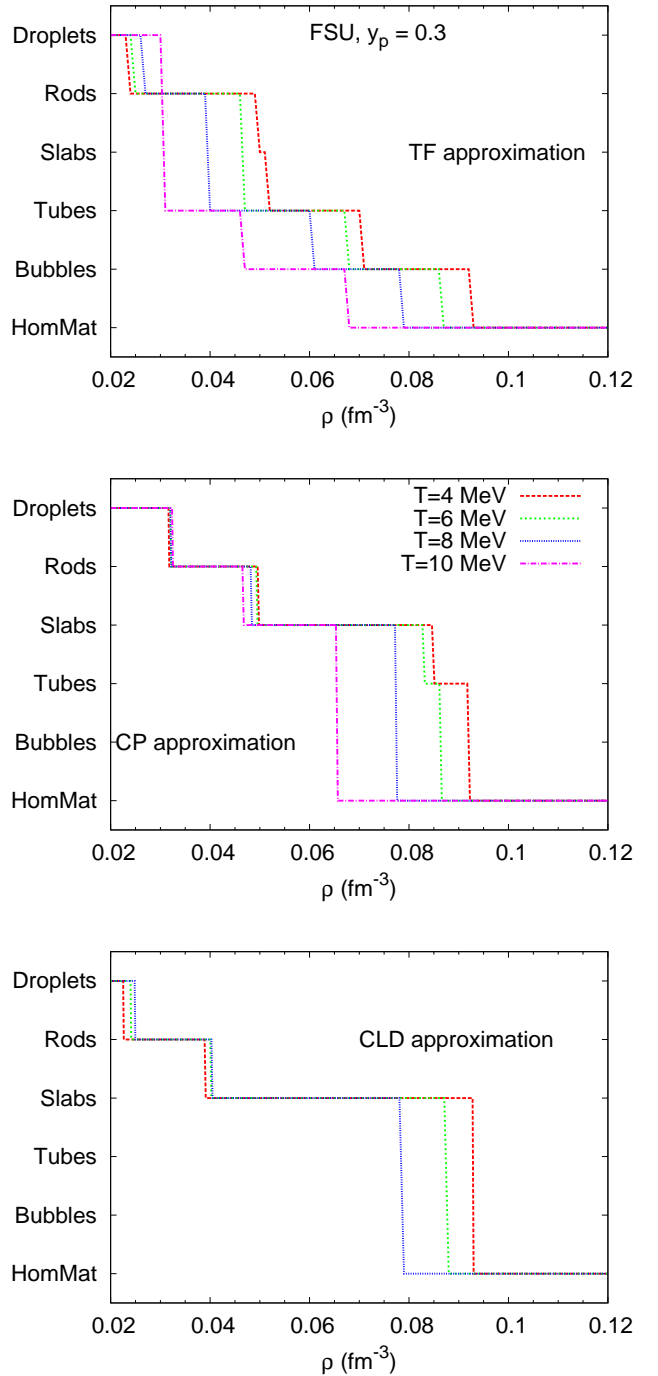}
   \end{tabular}
\caption{(Color online) Pasta phases for the FSU interaction within the TF (top), CP (middle) and CLD (bottom) calculations for several temperatures. The proton fraction is fixed to 0.3.}%
\label{fig12}
\end{figure}

In Ref.~\citep{Zhang14}, even though only droplets and bubbles were considered for simplicity, the transition densities to homogeneous matter are very close to our results with $T = 10$ MeV (see e.g. Fig. 1 of Ref.~\citep{Zhang14}).

Let us now compare how the different approaches describe the
transition between the different shapes. In Table \ref{tab2}, we give
the densities at the  droplet-rod, rod-slab, slab-tube, tube-bubble transitions, 
and the onset density for homogeneous matter.

These transition densities have been plotted in Fig.~\ref{fig12}, where the density range for each geometric configuration is shown for the TF (top panel), CP (middle panel) and
CLD (bottom panel) calculations, and a range of temperatures 4 to 10
MeV. It is interesting to notice that in the  CLD approach, tubes and
bubbles do not exist, and at $T = 10$ MeV, the pasta  geometries no longer exist.  For the CP calculation, the tubes and bubbles are also not
favored:  droplets, rods and slabs are present for all temperatures, but tubes only
briefly appear, at $T = 4$ and 6 MeV, and no bubble configuration was
found within this calculation. For the TF approximation, we found all
the shapes, except for the slabs, that only appear at $T = 4$
MeV.

\begin{figure}[!htbp]
   \begin{tabular}{c}
\includegraphics[width=0.5\textwidth]{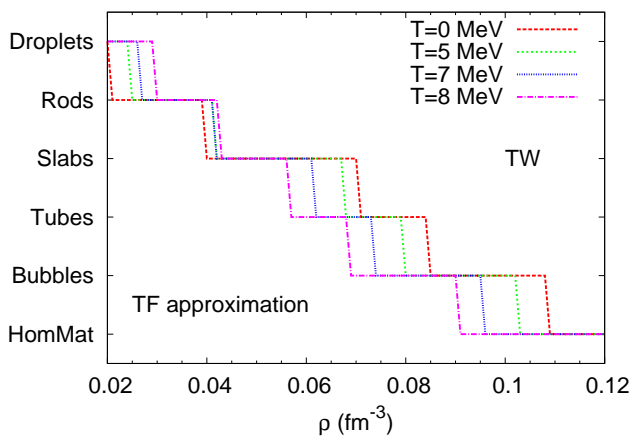}\\
   \end{tabular}
\caption{Pasta phases for the TW interaction within the TF calculation for several temperatures. The proton fraction is fixed to 0.3. (data taken from \cite{Avancini-12}).}   
\label{fig13}
\end{figure}

However, it is important to stress that the appearance and
density range of the different geometries  is model dependent, and
sensitive to properties such as the symmetry energy. 
In particular, in Figure \ref{fig13}, where the different pasta geometries are represented for the density dependent RMF model TW \cite{Typel-99}, for a proton
fraction of 0.3, we can see that all the five geometrical configurations are present
until $T=8$ MeV. The transition densities to uniform matter, as we already saw, are represented in Fig.~\ref{fig11}. In Ref.~\cite{Ducoin-08}, it has been shown that the
behavior of the symmetry energy, and its derivatives with the density, for
relativistic nuclear models with density dependent couplings, is much
closer to the behavior of Skyrme forces.
 
In Ref.~\cite{Maruyama-05}, it was discussed the
influence of a correct treatment of the Coulomb interaction on the
extension of each pasta geometry. In particular, in a calculation excluding the Coulomb  field, and including the Coulomb energy, as was done in the CP calculation, the bubble geometry was not
present with $Y_p=0.1$, and the slab configuration was found in a wider density range with $Y_p=0.3$. The same authors have also discussed the role of the surface tension, and have shown that a smaller surface tension favors a larger variety of geometries. In our CP and CLD calculations, the surface tensions have been calculated within the
model, however the treatment is not selfconsistent, and for the
larger densities, when the background nucleon gas becomes denser, the
surface energy used is probably too high and the neutron skin should have been included explicitly in the surface energy, as in Ref.~\cite{Douchin00}. This also explains why the
crust-core transition density calculated within the CP approach is smaller than the TF result.

\section{Conclusions} \label{IV}

In this work, we have studied the pasta geometries that appear in
core-collapse supernova events, and in the inner crust of neutron
stars, within three different approximations, the Thomas-Fermi, the
Coexisting Phases and the Compressible Liquid Drop calculations, all
within the single nucleus approximation. While
the first is a selfconsistent calculation, where the Coulomb interaction
and surface energy are adequately described, the other two
approaches are non-selfconsistent. They use as an input the surface energy for
semi-infinite matter as a function of the proton fraction and
temperature, obtained within a Thomas Fermi calculation. In the
CLD approach, the equilibrium conditions between the liquid and gas
phases take into account the Coulomb energy and surface
contributions, contrary to the CP method. An improvement of the last
two methods is tightly related with a realistic description of the
surface of the clusters.

We have introduced light clusters into our system to understand their
effect on the EoS. We observed that their effect is
only noticeable at very low densities, before melting.  It was shown
that taking light clusters into account always lowers the free energy. The
inclusion of light clusters in all the methods used in the present work to
describe the inner-crust will allow going beyond the single nucleus
approximation.

We were also interested in characterizing the transition to uniform matter. For that effect, we have plotted the free energy, pressure, entropy and chemical potentials to observe if there were any discontinuities. We realized that the density range of the pasta phase and the crust-core transition density decrease with increasing temperature, as expected, however the melting temperature of the different pasta phase geometries depends on the model properties. The stable geometries that we found depend on the parametrizations used, and the properties that influence them should be investigated. Also, the jumps in the pressure and chemical potentials, as a function of the density, could indicate a first order phase transition to uniform matter. Within the CP method, the description of a non-homogeneous phase gives unrealistic results at low densities, though it predicts concordant transition densities to uniform matter.   All the methods considered in this study show a very good agreement with respect to the transition density to homogeneous matter, and, in particular, the TF and CLD calculations give very similar results in the whole range of densities and temperatures considered.

\section*{ACKNOWLEDGMENTS}

Partial support comes from ``NewCompStar'', COST Action MP1304. H.P. and S.C. are supported by FCT under Projects No. SFRH/BPD/95566/2013, and SFRH/BPD/64405/2009, respectively. This work is partly supported by the project PEst-OE/FIS/UI0405/2014 developed under the initiative QREN. The authors acknowledge the Laboratory for Advanced Computing at the University of Coimbra for providing CPU time with the Milipeia cluster.

\end{document}